# The Cost and Benefits of Static Analysis During Development

## Quantitative Observational Results from Industry

William R. Nichols, Jr.
Software Engineering Institute
Pittsburgh, PA USA
wrn@sei.cmu.edu

## ABSTRACT

Without quantitative data, deciding whether and how to use static analysis in a development workflow is a matter of expert opinion and guesswork rather than an engineering trade-off. Moreover, relevant data collected under real-world conditions is scarce. Important but unknown quantitative parameters include, but are not limited to, the effort to apply the techniques, the effectiveness of removing defects, where in the workflow the analysis should be applied, and how static analysis interacts with other quality techniques.

This study examined the detailed development process data 35 industrial development projects that included static analysis and that were also instrumented with the Team Software Process. We collected data project plans, logs of effort, defect, and size and post mortem reports and analyzed performance of their development activities to populate a parameterized performance model. We compared effort and defect levels with and without static analysis using a planning model that includes feedback for defect removal effectiveness and fix effort.

We found evidence that using each tool developers found and removed defects at a higher rate than alternative removal techniques. Moreover, the early and inexpensive removal reduced not only final defect density but also total development effort.

The contributions of this paper include real-world benchmarks of process data from projects using static analysis tools, a demonstration of a cost-effectiveness analysis using this data, and a recommendation these tools were consistently cost effective operationally.

## KEYWORDS

## 1 Introduction

The economic challenges associated with fielding highly secure and cyber-resilient systems are well known and relevant to both industry and government[4][38]. A variety of tools and techniques are available for which specific strengths discussed at length in a State of the Art Report. [45]. A challenge comes when making cost benefit trade-offs select and apply these tools. Despite a variety of models that are intended to address cybersecurity planning and implementation ([23], [6], [17], [13], and [5]), the practical questions regarding the costs and benefits of using SSA techniques are little understood.

**The State of the Art Report,** [45] quoted Larsen on the problem from a DoD perspective, stating, "*There is a general lack of relevant quantitative data about the true costs, schedule impact, and effectiveness (in various situations) of specific tools, specific techniques, and types of tools/techniques […]. This lack of quantitative data makes selecting tool/technique types, and selecting specific tools, much more difficult,*"

In the absence of guidance in the form of data or validated models, the selection (software security assurance) SSA techniques and tools and how much effort to apply is a combination of expert judgement and guess work. Developing reliable, secure, and cyber-resilient software requires multiple software security assurance (SSA) interventions throughout the development lifecycle. These interventions include manual methods (e.g., reviews and inspections) as well as automated methods (e.g., static analysis and dynamic analysis).

This contribution of the research is the presentation of relevant in process data gathered from several static analysis tools in use under real world conditidtions, and demonstrating the cost effectiveness of these tools even limited to the scope of development. We report quantitative in process results that include static analysis from 35 projects in three orgzizations. The data is then used in quantitative models to estimate the costs benefit trade-off during the development activities. This report concludes with some lessons learned and the transferability to other contexts.

## 2 Background

Research on real-world issues includes a tradeoff between the level of researcher control and realism. This is essentially a tradeoff between internal validity and external validity. This tradeoff along with other related issues, such as compliance, coherence, and concordance, have been discussed in medical literature [30]. A distinctions is in the medical literature is made between effectiveness and efficacy ([37], [14], and [44]). This stylized semantic distinction is helpful to mind the gap between theory and implementation.

The efficacy of source code static analysis has a substantial history of research. Early work this century examining static analysis tool efficacy included a comparison of industry tools [11], an examination of fault detection [47], and a systematic literature review [15]. Austin studied of the effectiveness in detecting vulnerabilities[2] and recommended defense in depth [3]. Nagappan [25] reported a study at Microsoft for using static analysis to predict release defect density and, more recently, a study was performed on Chromium [24].

The effect of static analysis on developer behavior has been investigated at Google ([33] and [21]) and on effectiveness supplementing code review [29]. Alglave [1] reported a spectrum of practical issues involving the use of verification tools in general. Heckman [16] and Kim [20] investigated the prioritization of warnings.

While studies from [25] Microsoft and on Chromium [24] hint at effectiveness in that they investigate real systems, they either applied tools with high expertise, applied them outside the development process, or surveyed attitudes and beliefs rather than observing the behaviors and performance of developers. In none of these cases was actual development effort or defect containment tracked.

An unpublished source, Keane [39], reported substantial economic benefits from introducing static analysis, including reduced defect escapes with reduced total effort. Unfortunately, though plausible, the raw data is not available, and the work was not peer reviewed. The results are correlational observations from several overall project costs, so causality cannot be inferred. A more fine-grained measurement program would make the case more compelling.

While many aspects of static analysis and efficacy have been studied with rigor, a gap remains in understanding real-world effectiveness, especially the operational costs and benefits when applied by normal developers under real-world conditions.

## 3 Approach

### 3.1 Project Selection

The main barrier to this research was the ability (or willingness) of teams to gather the required data. Our criteria for participating this research included the following: (1) The projects must provide complete and detailed data comparable to that used by the Team Software Process (TSP). (2) The projects must have completed by the end of our funding year, 2017. (3) The data must be submitted to the SEI for use in our research. (4) Participants in the development projects must be available to provide context and explanation where necessary. (5) The project must include use of at least one static analysis tool in development in a way that could be identifiable in the data. (6) Multiple projects from the same organization should be available. The threats to validity with each of these criteria will be described in a later section.

We selected 35 projects from three organizations for analysis. A brief overview of the project types is provided in Table 1.

**Table 1: Static Analysis Tools and the Process Phases Where They Were Used**

| Org. | Number of Projects | Domain | When Tool Used |
|---|---|---|---|
| A | 5 | Avionics | Personal Review |
| B | 16 | Business Intelligence | Compile, Code Inspection, Personal Review |
| C | 14 | Design Automation | Build and Integration Test |

Of the 35 projects, 30 provided size data, additional project information from these projects provided in **Table** 2.

**Table 2: Development Team Information**

| Org | Team Size | A&M LOC | Hours | Days |
|---|---|---|---|---|
| A | 35 | 117279 | 7130.8 | 490 |
| A | 30 | 246118 | 7746.8 | 391 |
| A | 48 | 796887 | 35091.5 | 2215 |
| A | 41 | 84316 | 10851.6 | 457 |
| A | 36 | 89127 | 10927.9 | 490 |
| B | 16 | 20318 | 2626.3 | 246 |
| B | 8 | 37123 | 3950.7 | 552 |
| B | 13 | 484 | 537.4 | 47 |
| B | 13 | 1865 | 707.9 | 47 |
| B | 11 | 22411 | 1929 | 327 |
| B | 12 | 4020 | 1278.4 | 88 |
| B | 7 | 4494 | 924.5 | 53 |
| B | 15 | 6089 | 2248.8 | 844 |
| B | 10 | 442 | 788.5 | 92 |
| B | 9 | 5148 | 1234 | 264 |
| B | 7 | 38302 | 3165.5 | 272 |
| C | 3 | 23 | 21.2 | 21 |
| C | 19 | 1817 | 2294.6 | 149 |
| C | 19 | 737 | 572.2 | 374 |
| C | 5 | 4042 | 815.8 | 178 |
| C | 20 | 83 | 1296.8 | 129 |
| C | 9 | 8998 | 1941.9 | 140 |
| C | 22 | 2554 | 3207.8 | 512 |
| C | 8 | 9741 | 1282.5 | 73 |
| C | 25 | 46694 | 3392.4 | 137 |
| C | 8 | 9678 | 1282.8 | 91 |
| C | 8 | 7806 | 1141.9 | 72 |
| C | 7 | 13333 | 1532.2 | 114 |
| C | 25 | 66499 | 3294.8 | 180 |
| C | 11 | 6500 | 2253.7 | 316 |

Organization A used a systems engineering development workflow with  in which the projects were typically longer in cycle duration and larger in size and effort than the others.  The products were used internally. Organization B provided tools used in sales and marketing, and therefore focused on enhancements delivered duing specific yearly time frames. Organization C provided ongoing enhancements to a commercial product targeted for industrial use.

### 3.2 Static Analysis Tools Used

Each organization used a different tool of its own selection or in its own way. We regret that we cannot be more specific, but SEI institutional policy requires a written concent for the use of commercial product names that has not yet been obtained. The tools are identified only as A, B_1, B_2, and C, corresponding to the organization that used that tool. A brief description of the tool and its use is provided in the following text.

*Static Analysis Tool A* is a commercial static code analysis tool used to identify security, safety, reliability issues and complexity trends in C, C++, Java, and C# code. This tool finds memory leaks, vulnerabilities, and boundary conditions through techniques including build comprehension, code compilaton data flow and symbolic execution. This tool etool was typically used by developers prior to performing a personal review or peer review, however there was no overall usage policy.

*Tool_B_1* enforces a common set of style rules for C# code using a single, consistent set of rules, with minimal rule configuration allowed. Developers can implement their own rules if they so choose. This tool was used before personal code review.

*Tool_B_2* is typically used with tool Tool_B_1.  This tool analyzes managed code assemblies for the .NET Framework and reports information about the assemblies, such as possible design, localization, performance, and security improvements. This tool was used by developers during their desktop builds.

*Tool_C* is a commercial software development product consisting of both static code analysis and static binary analysis. It enables engineers to find defects and security vulnerabilities in source code written in C, C++, Java, C#, and JavaScript. using bug pattern matching, inter procedural data flow, abstract interpretation, false path pruning, Boolean satisfiability, design pattern intelligence, change impact analysis The tool was not directly visible to developers but was part of the automated build and integration system executed after code commit and before the system test.

### 3.3 Data and Workflow Description

All development teams produced code using an incremental approach in which discrete pieces of work passed through the logical stages of development. The specific sequence of activities varied both between and within organizations. In general, the creative activities began with some sort of requirements development or refinement, followed by high or detailed design, then code and integration. We categorize the types of work as product Creation, Appraisal, and Failure.

Creation activities (requirements elicitation, design, code) would normally be followed by a defect-removal activity such as a review, inspection, automated analysis, (Appraisal) or test (Failure). This separation of creation and defect-removal activities helps to separate work from rework with minimal disruption in workflow and is ideally suited for this type of study. The specific workflow activities are included in Table 5, Table 6, and Table 7.

The directly measured data we used, size, effort, and defects, are summarized in Table 3. Additional data were recorded and analyzed as part of a larger effort [26] that is beyond the scope of this study.

**Table 3: Directly Measured Data**

| Data | Description | Unit |
|---|---|---|
| Activity | The Type of Work Performed | Category |
| Defect Finds | Count of Defects Removed During an Activity | Count |
| Defect Fix-Time | Effort Required to Fix a Defect | Minutes |
| Effort | Effort Applied in an Activity | Minutes |
| Size | Lines of Code | Count |

The direct data were transformed into derived data for use in the modeling. These included activity rates for the creation of the product, rates of defect injection, fraction of defects removed, and the average time required to identify and fix a defect during an activity. The derived factors are summarized in Table 4.

**Table 4: Derived Data Used in Model**

| Data | Definition | Unit |
|---|---|---|
| Phase Effort Rate | Fix Effort | LOC/Hr |
| Defect Injection Rate | | Defects/Hr |
| Activity Yield | (Defects Removed)/Defects Present | None |
| Fix Cost | Reciprocal of Average Defect Fix Time | Hr/Defect |
| Size | Lines of Code | Count |

### 3.4 Data Collection and Storage

Individual developers collected the detailed effort, defect, and size data as they performed their work using the Process Dashboard tool [40]. The Dashboard file, containing all project data, was contributed to the SEI for research purposes, at a minimum at project completion, along with project postmortem analysis.

Most projects also contributed project launch Dashboards along with other launch artifacts. The Dashboard files were imported into the SEMPR Repository [35] using the Process Dashboard Data Warehouse [41]. Data queries were constructed to extract data into summary fact sheets [36], for analysis. These fact sheets, with identifiable information redacted, are available to other researchers at Carnegie Mellon Kilthub [28].

### 3.5 Modeling

Although useful for planning similar projects, the data require additional interpretation to evaluate effectiveness. TSP employs a software process improvement (SPI) approach with mechanisms to identify process areas for improvement, estimate the effects of change, and measure the actual results. Because of confounding factors and natural variation, simple pre-post designs can be hard to interpret or even impractical to interpret.

Instead, TSP teams typically use standard models calibrated with local data to examine each activity’s inputs and outputs. A removal activity, such as static analysis, would be modeled based on its cost to implement (in effort hours) and its effectiveness in defect removal. The TSP model is very similar to that used in COQUALMO ([22] and [8]) and by Raffo [32] and was further described by Nichols [27]. A brief description follows.

The TSP planning model for effort and defect containment is mathematically descriptive in that it will exactly match historical data. This model is deterministic, not stochastic. The assumed mechanism can be imagined using a tank and filter metaphor. Product creation (designs, source code, test cases, and so forth) fills the tank with not only product (requirements, code, designs, test cases) but also defects. Both product size and defects are modeled as a linear function of effort.

Defect removal requires additional steps following creation such as peer reviews, test, or static analysis to remove defects from the product. Defect removal activity effectiveness is measured using activity yield, which is the fraction of defects removed from the product by that activity. The effort in defect removal included the minimum time required to perform that activity if no defects were found, plus the time required to find and fix the defects detected. In this way, defects that escaped into test would require more effort to fix than those found in a code inspection. Likewise, manual execution of test always requires effort to both run tests and fix defects, while automated execution of test only requires effort to fix the defects found. We used activity average fix rates.

Essentially, each discrete piece of work requires some number of new and changed lines of code to implement. Each block of code takes time to design, code, review, or test. Each review or test activity finds some portion of the defects, and each defect requires some average effort to repair.

The model is calibrated using local historical parameters for the rates of production per hour, rate of defect injection per hour, defect removal yields, and average defect fix time by activity. The effects of reasonable changes can be estimated and propagated through the workflow to estimate their effect on effort and residual defect levels.

The model was initially developed to compare scenarios for which the best representations of the parameters and the outcome for the long term central tendency is a mean. While we report some project specific data and provide all data needed to perform the analysis on a project by project basis, we are more interested in the overall effect for an organization. Therefore, we chose to aggregated the and analyze the project parameter data by organization. The underlying data, however, support analysis of variation at a project level and, with some assumptions, at a component level. The variation at project level is left for future work.

## 4 Results

We briefly summarize some key results with respect to where defects were found, where effort was applied, and how the use of static analysis changes the effort and defect profile during the workflow. The quantitative data were collected for a related project [26] and the data have been made available for others to use [28].

### 4.1 Organization A

These five projects in the avionics domain were executed sequentially from system requirements through deployment. The commercial static analysis tool was used by the developers as code was written, around the time of their personal review of the code. In practice, the specific timing and useage varied but effort and defects were attributed to the personal code review (CODER) activity. Static analysis issues were identifiable from the defect log description field.

The organization average effort and defect-removal parametersfor the full development are summarized in Table 5. A notable feature of this data is that the **CodeReview** activity proceeded at an average rate of over 700 LOC/Hr with an average yield of 14%. In contrast, the peer review (CodeInspect) proceeded at a little over 200 LOC/Hr and achieved a find rate (yield) of 68%.

**Table 5: Organization A: Cost and Quality Performance Parameters**

| Phase | Phase Effort Rate LOC/Hr | Defect Injection Rate Def/Hr | Yield Without / WithSA | Fix Cost Hr/ Defect |
|---|---|---|---|---|
| BeforeDev | | | 0.00 | |
| Misc | 52.00 | 0.00 | | 0.42 |
| Strat | 5555.56 | 0.00 | | |
| Plan | 1470.59 | 0.02 | | 0.01 |
| SE_REQ | 40.98 | 0.05 | | 0.28 |
| SE_REQR | 230.41 | 0.03 | 0.25 | 0.07 |
| SE_REQI | 311.53 | 0.03 | | 0.05 |
| SE_REQval | | 0.01 | 0.12 | 0.48 |
| ReqReview | 20000.00 | 0.06 | | |
| ReqInspect | 245.10 | 0.00 | | |
| HLDesign | 243.90 | 0.08 | | 0.46 |
| ITPlan | | | | |
| HLDReview | | 2.09 | 0.01 | 0.07 |
| HLDInspect | 819.67 | 0.00 | 0.04 | 0.12 |
| DLDesign | 144.72 | 1.23 | | 0.35 |
| TestDevelop | 4347.83 | 0.26 | | 0.47 |
| DLDReview | 574.71 | 0.04 | 0.16 | 0.11 |
| DLDInspect | 300.30 | 0.04 | 0.60 | 0.10 |
| Code | 105.82 | 2.17 | | 0.32 |
| **CodeReview** | **709.22** | **0.04** | **0 / 0.14** | **0.09** |
| Compile | 2777.78 | 0.06 | 0.07 | 0.05 |
| CodeInspect | 213.68 | 0.06 | 0.68 | 0.13 |
| UTest | 178.25 | 0.02 | 0.51 | 0.44 |
| BITest | 333.33 | 0.05 | 0.40 | 0.51 |
| STest | | 0.00 | 0.17 | 3.94 |
| Doc | 429.18 | 0.04 | 1.00 | 0.15 |
| PM | 2564.10 | 0.00 | | |
| PLife | | 23.08 | 0.00 | 0.04 |

The results for defect removal with and without static analysis (starting with the code construction activites) are shown in Figure 1. The actual activity efforts are displayed in Figure 2.

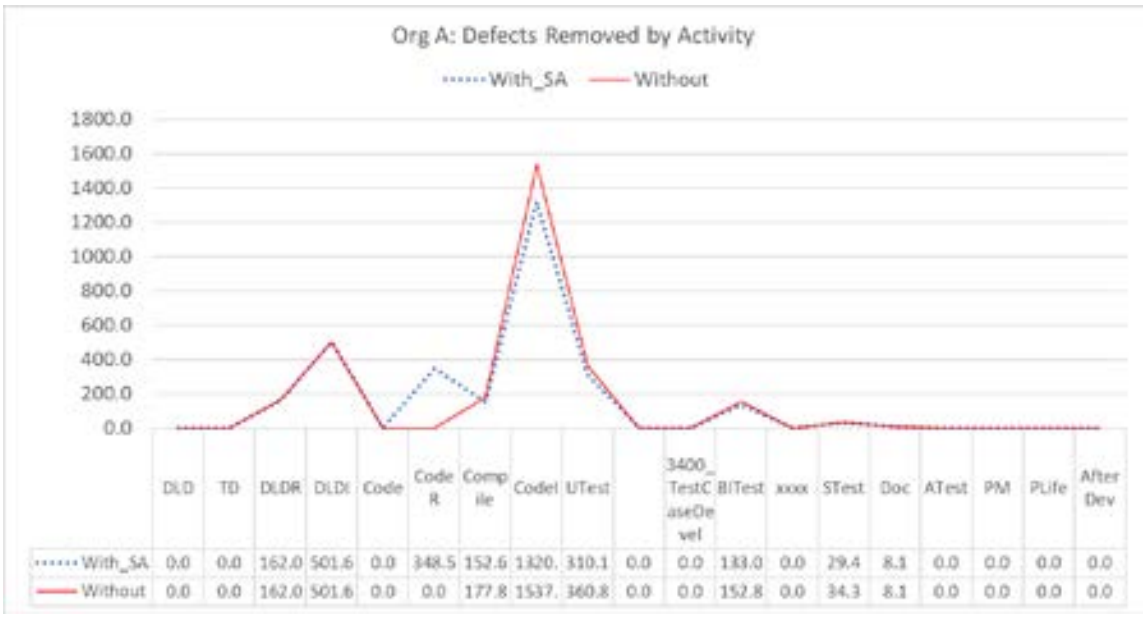

**Figure 1: Organization A: Defects Removed by Activity**

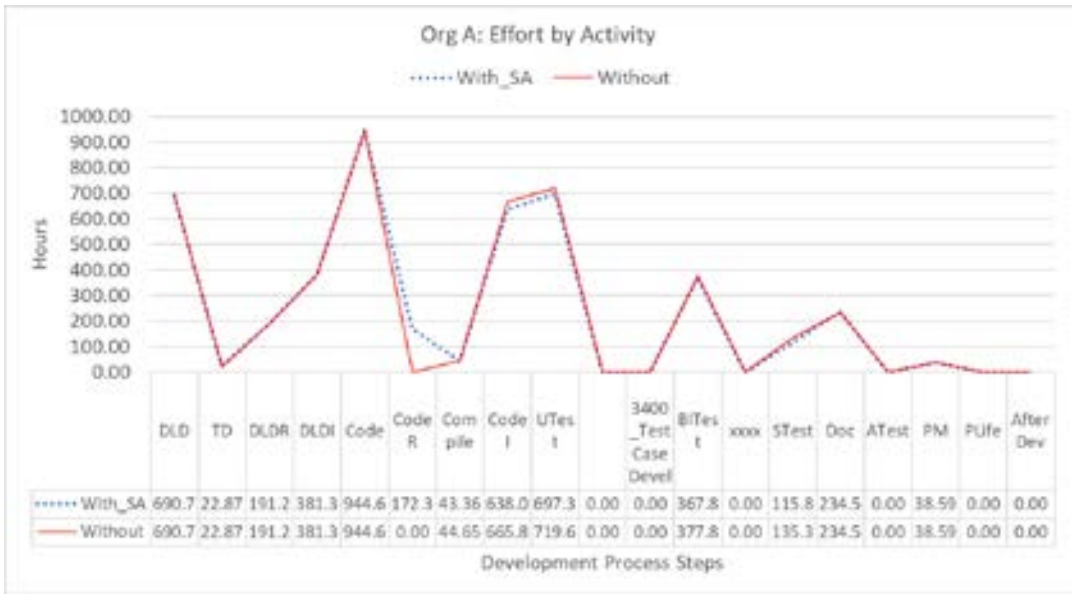

**Figure 2: Organization A: Effort by Activity**

## 4.2 Organization B

This organization used two related tools that were freely distributed with the development environment, one to examine source code and the other to check interfaces at build. Because tool use was not a separately measured activity, the counts and efforts are recovered from the defect logs.

The actual effort and defect removal parameters from the 16 projects in Organization B are summarized in Table 6. Removals occurred during the CodeR, Compile, and BITest phases. The defects found from the tools were counted separately and the model parameters were adjusted to lower yields in the hypothetical situation where the tools were not used between the build and system test activity.

**Table 6: Organization B: Cost and Quality Performance Parameters**

| Phase | Phase Effort Rate LOC/Hr | Defect Injection Rate Def/ Hr | Yield Without / WithSA | Fix Cost Hr/ Defect |
|---|---|---|---|---|
| Before Dev | | | 0.00 | |
| Misc | 128.05 | 0.00 | | |
| Strat | 903.55 | 0.00 | | |
| Plan | 164.22 | 0.03 | | 0.45 |
| Req | 209.05 | 1.39 | | 0.45 |
| ReqR | 612.80 | 0.01 | 0.21 | 0.04 |
| ReqI | 276.14 | 0.07 | 0.79 | 0.06 |
| HLD | 2238.85 | 1.49 | | 0.31 |
| ITP | 257.42 | 0.16 | 0.08 | 0.04 |
| HLDR | 13878.01 | 0.00 | 0.10 | 0.15 |
| HLDI | 1995.23 | 0.02 | 0.20 | 0.16 |
| DLD | 108.86 | 3.47 | | 0.22 |
| TD | 288.32 | 1.41 | | 0.18 |
| DLDR | 327.99 | 0.09 | 0.26 | 0.10 |
| DLDI | 86.20 | 0.11 | 0.48 | 0.12 |
| Code | 95.89 | 3.39 | | 0.16 |
| **CodeRev** | **332.35** | **0.10** | **024/ 0.25** | **0.09** |
| **Comp** | **1713.47** | **0.17** | **0 / 0.10** | **0.06** |
| **CodeInsp** | **64.43** | **0.08** | **0.63 / 0.65** | **0.10** |
| UTest | 160.24 | 0.06 | 0.55 | 0.31 |
| BITest | 122.46 | 0.05 | 0.22 | 0.14 |
| STest | 343.40 | 0.00 | 0.40 | 0.15 |
| AccptTest | 95.38 | 0.00 | 0.40 | 0.01 |
| PM | 1296.27 | 0.00 | | |
| PLife | 12385.47 | 0.00 | 0.40 | 4.35 |

The actual defect removals with and without static analysis are shown in Figure 3, along with the modeling estimate of defects removals had the static analysis activity not been performed. In this case, the difference in removals was small, but most noticeable during the compile activity.

The efforts with and without static analysis are displayed in Figure 4. The model projects very little difference in effort with or without this tool in the development environment, but the reduction in late defects costs slightly in test and product life.

Based on modeling and defect labeling, we estimated a company-average reduction of 11% in escapes, which is comparable to the lower bound of Organization A. The median review rate was 165 LOC/Hr for the projects with the most static analysis finds, while the median code review rate was 359 LOC/Hr for the others.

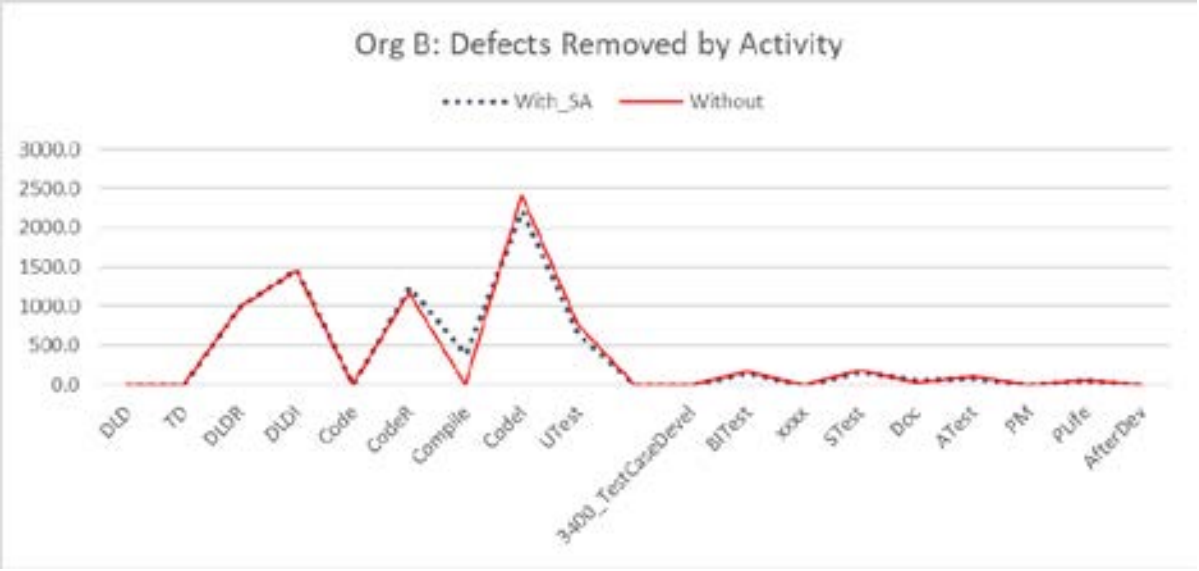

**Figure 3: Organization B: Defects Removed by Activity**

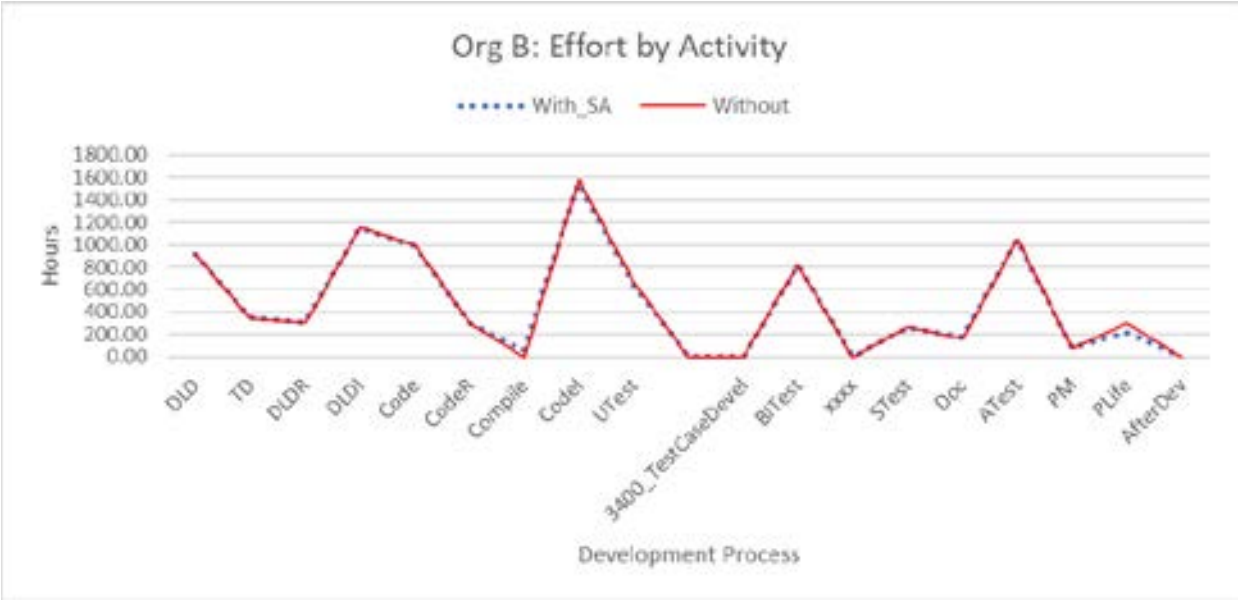

**Figure 4: Organization B: Effort by Activity**

### 4.3 Organization C

Organization C ran the tool during the build process, during which developers only logged effort to remediate the issues. Parameters are based on defect fix time and developer time logged duing that activity.

The effort and defect removal parameters from the 14 projects in Organization C are summarized in Table 7. Removals occurred between the build and system test activity. The static analysis examined both the source code and the binary code. The activity rate was a little under 400 LOC/Hr for a removal yield of 38%.

**Table 7: Organization C: Cost and Quality Performance Parameters**

| | Phase Effort Rate [LOC/Hr] | Defect Injection Rate [Def/Hr] | Yield Without / WithSA | Fix Rate [Hr/ Defect] |
|---|---|---|---|---|
| Req | 153.10 | 0.10 | | |
| ReqRev | 1184.20 | 0.00 | 0.03 | 0.04 |
| ReqInsp | 4439.30 | 0.00 | 0.27 | 0.03 |
| HLD | 580.70 | 0.10 | | 0.18 |
| ITP | 201426.20 | 25.20 | | 0.04 |
| HLDI | 50356.50 | 0.00 | 0.04 | 0.21 |
| DLD | 60.90 | 0.30 | | 0.64 |
| TD | 466.20 | 0.10 | | 0.67 |
| DLDRev | 247.70 | 0.00 | 0.19 | 0.19 |
| DLDInsp | 282.90 | 0.00 | 0.31 | 0.18 |
| Code | 32.20 | 0.40 | | 0.22 |
| CodeRev | 126.10 | 0.00 | 0.24 | 0.13 |
| Compile | 2047.50 | 0.10 | 0.05 | 0.02 |
| CodeInsp | 161.10 | 0.00 | 0.37 | 0.17 |
| Utest | 41.30 | 0.00 | 0.69 | 0.32 |
| BITest | 173.20 | 0.00 | 0.15 | 0.43 |
| **StaticAnalysis** | **392.40** | **0.00** | **0 / 0.38** | **0.22** |
| STest | 174.90 | 0.00 | 0.40 | 0.22 |
| PM | 1028.80 | 0.00 | | |
| PLife | 221.20 | 0.00 | 0.40 | 0.55 |

The results for defect removals are displayed in Figure 5, along with the modeling estimate of defect finds had the static analysis activity not been performed.

The activity efforts are displayed in Figure 6. The static analysis defect finds reduce defects in test and product life. The defect density after test changed from 1.9 to 1.2 Defects/KLOC.

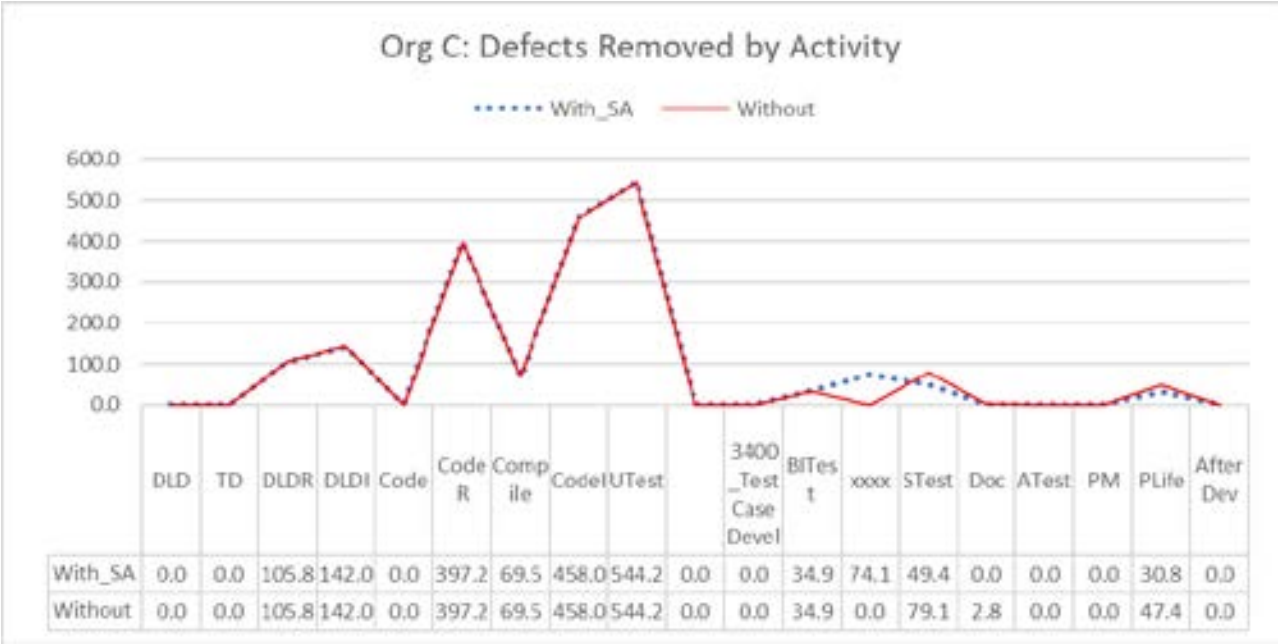

**Figure 5: Organization C: Defects Removed by Activity**

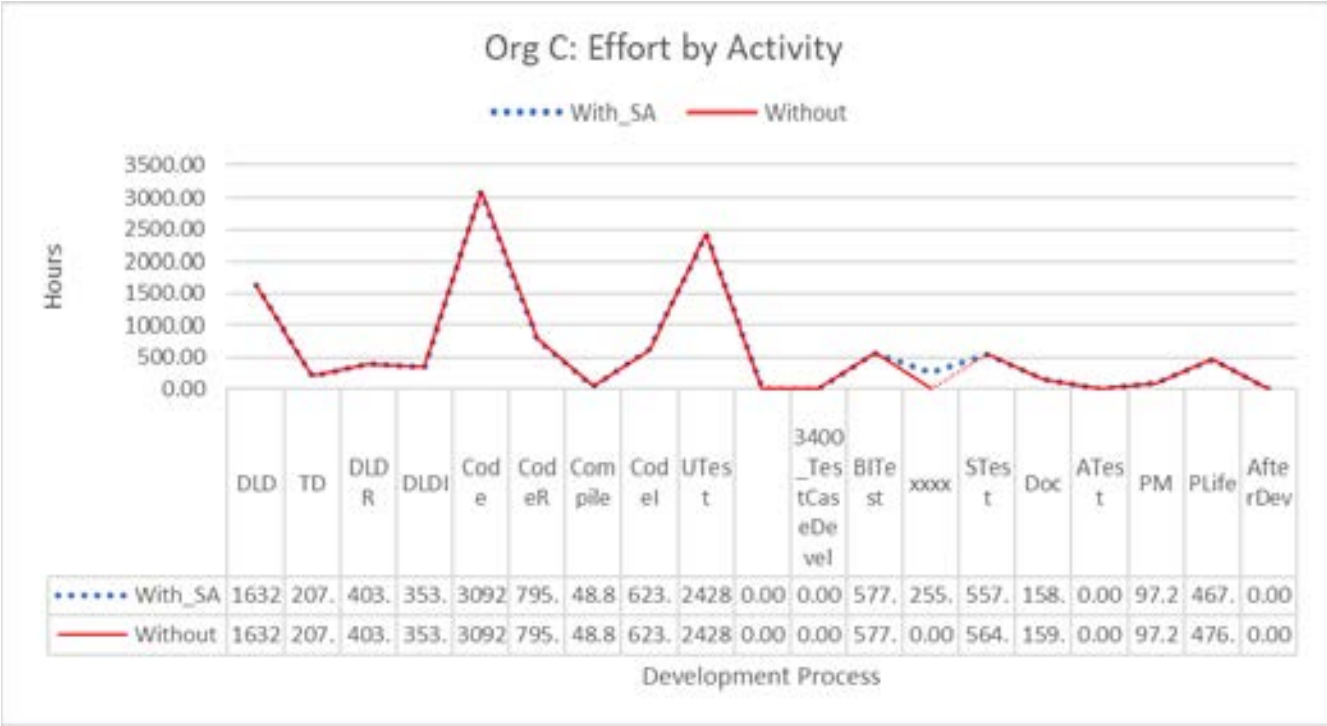

**Figure 6: Organization C: Effort by Activity**

## 5 Discussion

The specific challenges in this research are 1) to determine the operational costs associated with using these tools,2) measureing the benefits in schecule or effort, and 3) drawing conclusions about the generalizability or transferability of the results to other Contexts.  A and B, the tools were applied inconsistently among projects and developers. This inconsistency challenged evaluating effectiveness. Nonetheless we were able obtain sufficient data to assess these projects because we had access to detailed records of effort and defect descriptions. How to interpret the findings is the subject of the following sections.

### 5.1 How Effective Were Static Analysis Tools?

While the organizations employed different the static analysis tools in different ways, the overall results were consistent in that they enabled efficient finding and fixing of defects. Whenever static analysis was used it removed defects not previously found by other techniques. That is, regardless of the specific tool used, each tool found new defects with sufficient specificity to make the fix efficient. Moreover, the removed defects needed not be found by later, reducing effort in later bug fix activites. Essentially, the static analysis was cheaper operationally than finding and fixing defects in test.

Organization C, which automated code and static analysis at build time, had a small absolute improvement but substantial reduction (35%) in test failures. The low defect levels entering build resulted in few finds but a sizable yield. Despite an objectivel effective process (low defect levels entering build) this tool found defects that escaped earlier removal activities.

For organization A, the tool removal yields were not especially high (14%) nor for B, which had a 10% yield during compile and small increases in yield during the surrounding activites. Nonetheless, the removal rates were very high. The specificity of the warnings made fixes very fast. In each case, the fix times were faster than later phases by a modest amount. Both A and B, however, also benefited from very high yield inspections. Of course, all inspection defects had escaped static analysis and both organizations showed exceptionally high inspection yields. A common opinion expressed by developers was that the inspections could be more focused because the tools had removed certain types of defects, which could now be safely ignored.

The modest yields raised the question of how much overlap tools of the same type will actually have. One company that did not participate in the study indicated that they used multiple static code analysis tools to achieve higher overall yields. Though we were not able to verify this, the conjecture is consistent with both the low individual yields and the differing focus of the individual tools.

### 5.3 Lessons Learned

Each of the results from all three organizations using different types of static analysis tools leads to observations that 1) static analysis found bugs not discovered by other means, 2) the removal yields of a single tool were small, somewhere in the 15%-35% range, and 3) the find and fix rates were among the fastest of all removal techniques.

Each of the tools were cost effective operationally but this includes only a portion of the overall costs. Additional costs include tool licensing and training. This study is limited to the operational cost and effectiveness of applying static analysis tools for the subject projects. Nonetheless, the results may prove useful in making economic decisions for tool acquisition and training.

Obtaining measures was a challenge. No developers or project managers were willing to commit additional resources to adding measurement or automating tool use for this study. Moreover, none of these three organizations had performed a detailed economic analysis of the cost effectiveness with the available data we used. It is unclear how projects not instrumented like those in our study could make an objective determination. This will be especially challenging when considering if and how to combine multiple tools into the workflow.

Isolating tool usage from other activites was helpful with the accounting for defect yield and defect fix rate. Organization C had moved the furthest along this path, but collected no additional data. This is likely the simplest path for projects that want to measure the tool costs.

Given the modest effects of invidual tools, the cost/benefit tradeoffs may not be apparent to management or software acquirers. We have only anecdotes on why organizations introduced static analysis, most hoped to reduce defect escapes but none were based the decision on a micro-economic cost/benefit analysis. The data in this report provides some benchmarks. Anecdotal resistance, however, usually involved developers claiming the tools produced too many warnings and slowed them down.

Transferring these results into other contexts depends largely upon the defect fix rates and yields.Each of these organizations had effective defect removal that resulted in fairily low defect escapes and only small differences in fix rates. The models suggest that higher escapes and/or slower fix rates in later activities will increase the cost benefit of the tools. The benefits we estimated in this report, though modest, are also likely to be conservative. Other questions of validity are addressed in the next section.

## 6 Limitations and Threats to Validity

This study is limited to condering only the operational cost and effectiveness of applying static analysis tools for the subject projects. A more complete treatment of cost/benefit would include acquisition and licensing costs of the tools, the cost of fielded defects, and external costs, such as impact on the user. Given additional cost information, a more complete treatment can be found in Zhen. [47]. This study, however, can set expectations for performance improvements.

While some of the findings may generalize, the tools have been used in idiosyncratic ways. Moreover, because performance of specific tools can change from release to release [7], the performance of the tools in this study may change over time.

While this study has shown that static analyses of code and binary are finding unique defects when compared to reviews and test, we have very limited data on overlapping coverage from different tools of the same type, or with other of the many types of tools available. Much research remains in this area. This study’s contribution is to point toward some practical measurement practices to aid future work.

This study does not address training, user perceptions of the tools, or the actual gap between tool capability and practical use. Though we present some evidence that may indicate effects on behavior, we did not directly address user wants, needs, or opinions. Nor did we attempt to determine what settings were being used by the developers or organizations. How the tools could be used more effectively is a matter for future work and is beyond the scope of this study. We do not attempt to measure, false positives, nor the effect of false positives on the work. Instead, we measure the total effort applied and the work remediating those declared to be true positives.

This study did not attempt to demonstrate a direct correlation between static analysis finds and the prevention of faults. We directly measured the finds that were remediated and estimated the effects based on historical downstream find rates.

We do not evaluate the severity of the items found, which could be considered another measure of effectiveness. Such measures are beyond the current scope of these development teams but might be added to future work by capturing information, such as CWE references in the logs. The participants were unwilling to add to their work scope for this study.

In addition to the limitations, it is important to understand the threats to validity. We briefly discuss some external, construct, and internal validity issues in this work.

### 6.1 External Validity

It is important to appreciate the threats posed by the project selection criteria in section 3.1. That the projects have completed seems a minor risk unless we believe that static analysis leads to more failures and we have a survivor bias.

Several criteria will tend to select for organizations that are larger, more capable, or better resourced organizations. These include the requirement of multiple projects and the use of TSP or other sufficiently detailed data. In particular, TSP teams that record data and send it to SEI have already demonstrated a dedication to quality [18], process discipline and improvement. Projects with fewer resources were dropped from our study when they failed to implement the minimum required data collection requirements.

The high quality of these projects may also bias results in that processes with fewer defect-removal options may find enhanced benefits from including static analysis.

The capable projects that did not partipate also bias the results. For example, we could not include data from a medical devices company that had achieved exceptional quality results while employing static analysis tools because local privacy laws restricted release of their data.

Another potential source of bias was the requirement that the static analysis data be identifiable within the tool. This certainly reduced the sample, but we can only speculate about bias.

We cannot claim that results are representative for all kinds of software projects. Nonetheless, the results from these 35 projects included very different domains. Studying more projects would provide more confidence in the actual ranges found in practice. An effectiveness study gains strength with a broader sample.

Other domains will have different parameters, and some may have substantially higher static analysis yields, higher or lower costs of static analysis remediation, higher static analysis fixed costs (e.g., higher rates of false positives or difficulty disposing of them), or much lower costs of remediation in test (coupled with high test yields). Organizations A and C in particular operate in an environment in which high quality is essential because defects can have life threatening or high economic costs. Comparable industries would include embedded systems and medical devices. Organization B, business intelligence, may be more comparable to information or banking systems.

It is possible, for example, that the difference in fix times for system test and code review is much different than our measured averages. Moreover, the defect injection rates with an organization were more consistent than the removal yields. Also, static analysis should consistently find certain types of defects if they are present, and this is not true of test. Nonetheless, consistency of defect injection suggests that there will be opportunities to remove them with some effective approach.

In summary, this study lacks the breadth to put external validity concerns to rest. Nonetheless, it contributes a sample of real-world results.

### 6.2 Construct Validity

Construct validity assesses whether the study findings could be incorrect because of misleading data or incorrect assumptions in the model.

TSP data quality has previously been studied [35], [34]. The projects that were used all contained data that passed basic required consistency checks, including distributional properties and consistency among the logs. The association of defect finds with the tool was discussed with the data analysts.

We make the simplifying assumption that all defects are removed with similar effectiveness in different phases and that the finds will be more or less random. We know different types of activities remove defect types at different rates [42], [43]. Addressing this threat remains for future work.

The model also uses average values for the organizations. Mathematically, the project parameters will reproduce, ex post, the project results. In practice, parameters vary both statistically and because of process drift. With sufficient data, the model could be extended to use distributions to produce probability ranges for the outcomes. This remains for future work.

While we measure the defect findings during development, we do not directly measure fielded faults, severity, or the association of the defects with security. We assume relationships found in prior work [46], [45], [9], [10].

### 6.3 Internal Validity

Internal validity evaluates whether causal findings are due to artifacts of the study design and execution. This may include factors that have not been controlled or measured, or some introducing some unintended factor. Causality, by design, is beyond the scope of this work because attempts to control the behavior would weaken external validity. Instead, we assume a causal system and attempt to measure the effects of the tools as implemented in real-world environments. Internal validity could suffer to the extent that the assumptions were violated by subjective judgments, or by the participants or the analyst.

In everyday real-world development settings, factors such as developer or management preferences, process compliance, education, training, and so forth will affect the results. Although some of these factors could be accounted for with additional observation or surveys, these were beyond the scope of this work.

Analyst bias in conducting the research is another potential threat. We have attempted to minimize this by relying on quantitative data, fully disclosing that data, and using well known models in the analysis.

## 7 Conclusions

We measured defect removal effectiveness (yield) and effort applied for static analysis on 35 projects across three organizations. We used precise data records at a very granular level of detail to gain insight into the cost effectiveness of using static analysis in industrial projects software development workflows. Phase containment and effort modeling suggested that applying static analysis not only reduces escaped defect density but also total development effort regardless of whether it wsas applied shortly after initial coding or during the build. The use of static analysis removed defects that had escaped other defect removal at a cost comparable to other defect removals during contemporaneous activities.

We also found that defects found by static binary checkers checkers had find/fix effort no worse than other integration and system test defects. Defects found durning development had comparable fix efforts to those found during inspections. Nonetheless, in the context of a complete development process including multiple defect removal activites, static analysis yield was typically lower that the review or test.. We do not know, however, if the lower yield is a result of tool limitations or the way they were used. Additional research into tool effectivess and ways to improve performance in use is needed.

Our observational evidence suggests that static analysis tools are not yet as effective as some traditional methods but that they are finding defects that escaped other removal methods. Because they provide incremental improvments, they should be used in addition to rather than a replacement for those other removals methods. That is, operationally static analysis provides a positive marginal benefit for the cost.

Finally, this research has provided a data collection and modeling approach that can be applied to researching many of the open questions that remain.

Ideally, the processes workflow should be composed of a sequence of activities that work together as an effective and efficient whole, but to accomplish that requires detailed understanding of the development process. To answer our sponsor’s implicit questions, operationally at least, integratging these tools into development offers a cost effective option for defense in depth.

## ACKNOWLEDGMENTS

The author thanks David Tuma and Tuma Solutions for donating the Team Process Data Warehouse software for our research using TSP data and the work of Yasutaka Shirai for the development of the Software Engineering Measured Performance Repository (SEMPR) at the SEI that made this analysis possible.

This material is based upon work funded and supported by the Department of Defense under Contract No. FA8702-15-D-0002 with Carnegie Mellon University for the operation of the Software Engineering Institute, a federally funded research and development center.